\documentclass[twocolumn,times]{aastex631}

\usepackage[utf8]{inputenc}
\usepackage{natbib}
\usepackage{graphicx}
\usepackage{soul}
\usepackage{amsmath}
\usepackage{booktabs}
\usepackage{longtable}
\usepackage{xspace}
\usepackage{hyperref}
\usepackage[T1]{fontenc}
\usepackage{lipsum}
\usepackage{comment}
\usepackage{xcolor}
\usepackage{multirow}
\usepackage{IEEEtrantools}
\usepackage{amsmath}

\usepackage[super]{nth}

\newcommand{\teff}{\ensuremath{T_{\rm eff}}\,}

\newcommand{\logg}{\ensuremath{\log g_*}\,}

\newcommand{\msun}{\ensuremath{\,M_\Sun}}
\newcommand{\rsun}{\ensuremath{\,R_\Sun}}

\newcommand{\gaia}{{\it Gaia} }

\newcommand{\degree}{\ensuremath{\,^{\circ}}}

\begin{document}

\title{Evidence for Primordial Alignment II: Insights from Stellar Obliquity Measurements for Hot Jupiters in Compact Multiplanet Systems}

\author[0000-0002-0015-382X]{Brandon T. Radzom}
\affiliation{Department of Astronomy, Indiana University, 727 East 3rd Street, Bloomington, IN 47405-7105, USA}

\author[0000-0002-3610-6953]{Jiayin Dong}
\altaffiliation{Flatiron Research Fellow}
\affiliation{Center for Computational Astrophysics, Flatiron Institute, 162 Fifth Avenue, New York, NY 10010, USA}
\affil{Department of Astronomy, University of Illinois at Urbana-Champaign, Urbana, IL 61801, USA}

\author[0000-0002-7670-670X]{Malena Rice}
\affiliation{Department of Astronomy, Yale University, 219 Prospect St., New Haven, CT 06511, USA}

\author[0000-0002-0376-6365]{Xian-Yu Wang}
\affiliation{Department of Astronomy, Indiana University, 727 East 3rd Street, Bloomington, IN 47405-7105, USA}

\author[0000-0002-8685-5397]{Kyle Hixenbaugh}
\affiliation{Department of Astronomy, Indiana University, 727 East 3rd Street, Bloomington, IN 47405-7105, USA}

\author[0000-0002-4891-3517]{George Zhou} 
\affiliation{University of Southern Queensland, Centre for Astrophysics, West Street, Toowoomba, QLD 4350 Australia}

\author[0000-0003-0918-7484]{Chelsea X. Huang}
\affil{University of Southern Queensland, Centre for Astrophysics, West Street, Toowoomba, QLD 4350 Australia}

\author[0000-0002-7846-6981]{Songhu Wang}
\affiliation{Department of Astronomy, Indiana University, 727 East 3rd Street, Bloomington, IN 47405-7105, USA}


\newcommand{\tic}{TIC-281837575\xspace}
\newcommand{\toi}{TOI-5143\xspace}
\newcommand{\toib}{TOI-5143b\xspace}
\newcommand{\toic}{TOI-5143c\xspace}

\newcommand{\ra}{11:01:27.63}
\newcommand{\dec}{+05:08:22.13}
\newcommand{\parallax}{$5.5942\pm0.0189$}
\newcommand{\Gmag}{$11.8203\pm0.0008$}
\newcommand{\GBPmag}{$12.2318\pm0.0034$}
\newcommand{\GRPmag}{$11.1964\pm0.0019$}

\newcommand{\thismstar}{$0.864^{+0.040}_{-0.033}$}
\newcommand{\thisrstar}{$0.852\pm0.020$}
\newcommand{\thisrhostar}{$1.97^{+0.18}_{-0.16}$}
\newcommand{\thislogg}{$4.514^{+0.029}_{-0.027}$}
\newcommand{\thisteff}{$5183\pm125$}
\newcommand{\thisfeh}{$0.107\pm0.044$}
\newcommand{\thisage}{$7.4^{+4.4}_{-4.2}$}
\newcommand{\ttvper}{5.2097117$\pm 0.0000036$}
\newcommand{\ttvmidt}{2782.51853$^{+0.00023}_{-0.00024}$}

\newcommand{\DTrprs}{0.175$^{+0.025}_{-0.010}$}
\newcommand{\DTrhostar}{2.07$^{+0.13}_{-0.16}$}
\newcommand{\DTvsini}{2.53$^{+0.28}_{-0.29}$}
\newcommand{\DTb}{0.98$\pm 0.03$}
\newcommand{\DTua}{0.63$^{+0.32}_{-0.50}$}
\newcommand{\DTub}{0.13$^{+0.43}_{-0.46}$}
\newcommand{\DTnonrotv}{1.49$^{+0.20}_{-0.19}$}
\newcommand{\DTlam}{2.1$^{+2.8}_{-2.7}$}
\newcommand{\DTmidt}{2688.7485$^{+0.0010}_{-0.0010}$}

\newcommand{\DTperiodc}{5.2097118$^{+0.0000032}_{-0.0000039}$}
\newcommand{\DTmidtc}{2527.24264$^{+0.00030}_{-0.00027}$}

\newcommand{\DTrp}{1.45$^{+0.18}_{-0.12}$}
\newcommand{\DTap}{0.05602$^{+0.00082}_{-0.00079}$}
\newcommand{\DTaor}{14.14$^{+0.40}_{-0.37}$}
\newcommand{\DTincl}{86.02$^{+0.16}_{-0.15}$}

\newcommand{\transitrprs}{0.163$\pm 0.022$}
\newcommand{\transitrhostar}{2.03$\pm 0.14$}
\newcommand{\transitb}{0.98$\pm 0.03$}
\newcommand{\transitua}{0.39$^{+0.17}_{-0.39}$}
\newcommand{\transitub}{0.16$^{+0.36}_{-0.37}$}
\newcommand{\transitmidtczero}{2527.24122$^{+0.00089}_{-0.00092}$}
\newcommand{\transitmidtcone}{2532.45115$^{+0.00089}_{-0.00088}$}
\newcommand{\transitmidtctwo}{2537.66269$^{+0.00087}_{-0.00093}$}
\newcommand{\transitmidtcthree}{2542.87331$^{+0.00096}_{-0.00092}$}
\newcommand{\transitmidtcfour}{2548.08096$^{+0.00086}_{-0.00086}$}
\newcommand{\transitmidtcfive}{2558.50089$^{+0.00081}_{-0.00087}$}
\newcommand{\transitmidtcsix}{2563.71231$^{+0.00081}_{-0.00076}$}
\newcommand{\transitmidtcseven}{2568.92023$^{+0.00083}_{-0.00087}$}
\newcommand{\transitmidtceight}{2574.12953$^{+0.00081}_{-0.00076}$}
\newcommand{\transitmidtcnine}{3267.02198$^{+0.00078}_{-0.00079}$}
\newcommand{\transitmidtcten}{3272.23155$^{+0.00085}_{-0.00084}$}
\newcommand{\transitmidtceleven}{3277.44167$^{+0.00089}_{-0.00079}$}
\newcommand{\transitmidtctwelve}{3282.64982$^{+0.00088}_{-0.00082}$}


\begin{abstract}

A significant fraction of hot Jupiters have orbital axes misaligned with their host stars' spin axes. The large stellar obliquities of these giants have long been considered potential signatures of high-eccentricity migration, which is expected to clear out any nearby planetary companions. This scenario requires that only isolated hot Jupiters be spin-orbit misaligned while those with nearby companions, which must have more quiescent histories, maintain low-obliquity orbits, assuming they formed aligned within their primordial protoplanetary disks. Investigations of this stellar obliquity-companionship connection, however, have been severely limited by the lack of hot Jupiters found in compact multi-planet systems. Here we present the sky-projected stellar obliquity ($\lambda$) of a hot Jupiter with a nearby inner companion recently discovered by NASA's Transiting Exoplanet Survey Satellite: TOI-5143\,c. Specifically, we utilize the Doppler shadow caused by the planet's transit, enabled by the Rossiter-McLaughlin (RM) effect, to find that the planet is aligned with $\lambda=2.1 ^{+2.8}_{-2.7} \degree$. Of the exoplanets with RM measurements, TOI-5143\,c becomes just the third hot Jupiter with a nearby companion, and is part of the 19th compact multi-planet single-star system, with an RM measurement. The spin-orbit alignment of these 19 systems provides strong support for primordial alignment, and thus implies that large obliquities are gained primarily due to post-disk dynamical interactions such as those inherent to high-eccentricity migration. As such, the observed spin-orbit alignment of hot Jupiters with nearby companions affirms that some fraction of these giants instead have quiescent origins.

\end{abstract}


\section{Introduction} \label{sec:intro}

The origins of hot Jupiters remain one of the most well-studied yet contentious questions in exoplanet science. These short-period ($P<10\,\mathrm{d}$) gas giants ($M_{\rm p}>0.3 \, M_\mathrm{J}$) have been subject to intense observational campaigns over the past three decades that have enabled substantial population-level characterization (see \citealt[][and references therein]{Dawson2018}). Critically, in contrast with our solar system's Jupiter, a significant fraction of hot Jupiters are observed to have large stellar obliquities --- that is, their orbital angular momentum normal axis is often misaligned with the spin axis of their host stars (e.g., \citealt{Hebrard2008,Schlaufman2010, Winn2010a, Albrecht2012}, see reviews by \citealt{Winn2015, Triaud2018,Albrecht2022}), and an even greater majority of hot Jupiters are found to be isolated, devoid of any nearby planetary companions \citep{Steffen2012, Huang2016, Hord2021, Wu2023}.

Together, these observational properties are suggestive of relatively violent dynamical histories for these systems. In this regard, perhaps the most plausible origin is high-eccentricity migration, wherein these giants once occupied wider orbits and acquired large eccentricities following the dispersal of their protoplanetary disks. Mechanisms such as Lidov-Kozai oscillations \citep{Wu2003,Fabrycky2007,Naoz2016}, planet-planet scattering \citep{Rasio1996, Chatterjee2008}, or secular chaos \citep{Wu2011, LithwickWu2014, Hamers2017} may have excited their eccentricities sufficiently to trigger inward migration and subsequently circularization through tidal interactions with their host star, clearing out any nearby planetary companions \citep{Mustill2015}. While the true prevalence of the various proposed spin-orbit misalignment mechanisms is disputed \citep{Albrecht_2021, Dong_2023b, Siegel_2023}, the above processes responsible for high-eccentricity migration may play a major role (see also \citealt{Wang2021, Rice2022, Albrecht2022}).

If this interpretation holds, spin-orbit misalignment for hot Jupiters should be primarily restricted to those that are isolated (i.e., without nearby/close-in planetary companions). Conversely, hot Jupiters that have nearby companions would be expected to maintain orbits in close alignment with the stellar equator, as the presence of these companions inherently precludes a dynamically violent history. The connection between stellar obliquity and the companionship rate of these giants, however, is poorly understood as the former population of isolated hot Jupiters is observationally abundant while the latter appears nearly non-existent--- until recently.

The first of such compact multi-planet systems to be confirmed was WASP-47, which contains a $P=4.2 \, \mathrm{d}$ hot Jupiter surrounded by an inner super-Earth at $P=0.79 \, \mathrm{d}$ and an outer hot Neptune at $P=9.0 \, \mathrm{d}$ \citep{Hellier2012,Becker2015}. Soon after the discovery of this system's compact configuration, \cite{Sanchis-Ojeda2015} followed up with radial velocity (RV) observations to measure its stellar obliquity via the Rossiter-McLaughlin (RM) effect \citep{Holt1893,Rossiter1924,McLaughlin1924,Queloz2000}, finding the system was spin-orbit aligned. The \emph{Kepler} space telescope, still responsible for the discovery of most confirmed planets today, found only one hot Jupiter in a compact multi-planet system (Kepler-730; \citealt{Canas2019, Zhu2018}), though the host star remains too faint for ground-based spectroscopic follow-up ($V=15.8$; \citealt{Everett2012}), including detections of the RM effect. Besides WASP-47, the only other compact hot Jupiter systems with RM measurements are WASP-84 (a hot Jupiter and an inner super-Earth, \citealt{Anderson2015, Maciejewski2023}) and WASP-148 (a borderline hot Jupiter or hot Saturn with a nearby warm Jupiter companion, \citealt{Hebrard2020, Wang2022, Knudstrup_2024}), both of which were found to be well aligned.

In light of this, the ongoing Transiting Exoplanets Survey Satellite (TESS) mission \citep{Ricker2015} has revolutionized the field, uncovering several close-in gas giants with nearby companions (e.g., TOI-2202; \citealt{Trifonov2021}, WASP-148; \citealt{Hebrard2020}, TOI-2000; \citealt{Sha2023}, TOI-5126; \citealt{Fairnington2023}, TOI-5398; \citealt{Mantovan2022, Mantovan2024a}, HIP 67522; \citealt{Barber2024}), including hot Jupiters TOI-1130 \citep{Huang2020}, WASP-132 \citep{Hord2022}, TOI-1408 \citep{Korth2024}, TOI-2494, and TOI-5143 \citep[Quinn et al. 2025, in preparation;][]{Guerrero2021}. Precise stellar obliquity constraints for these systems are possible (e.g., see \citealt{Heitzmann2021,Rice2023b,Mantovan2024b,Radzom_2024}) thanks to the relative brightness of their host stars and the current suite of Extreme Precision Radial Velocity (EPRV) instruments that enable RM measurements on slow rotating stars. Once observed, the RM effect can be modeled in several ways, including the Doppler shadow technique \citep{Albrecht2007, Collier2010a}, which relies on distortions in the spectral line profiles during transit, rather than shifts in the radial velocity derived from the observed lines.

In this work, we present a Doppler shadow RM measurement of the sky-projected obliquity ($\lambda$) for the compact multi-planet system TOI-5143, using the EPRV NEID spectrograph. 
TOI-5143 is a KV-type star ($V=11.9 \, \mathrm{mag}$) that hosts a 5.2-day hot Jupiter (planet c), recently confirmed by Quinn et al. (2025, in preparation), and a 2.4-day sub-Neptune (planet b).
We find $\lambda=2.1 ^{+2.8}_{-2.7} \degree$ for TOI-5143\,c, revealing that the hot Jupiter is spin-orbit aligned. This finding continues the trend of alignment seen for compact multi-planet single-star systems \citep{Albrecht2013, Wang2018, Zhou2018}, providing compelling evidence for primordial alignment and hence post-disk misalignment through high-eccentricity migration channels. Consequently, the growing sample of aligned compact hot Jupiter systems demonstrates that a subset of these giants have dynamically quiescent histories.

This is the \nth{14} published work from the Stellar Obliquities in Long-period Exoplanet Systems (SOLES) survey (\citealt{Rice2021}; \citealt{Wang2022}; \citealt{Rice2022b}; \citealt{Rice2023a}; \citealt{Hixenbaugh2023}; \citealt{Dong_2023a}; \citealt{Wright2023}; \citealt{Rice2023b}; \citealt{Lubin2023}; \citealt{Hu2024}; \citealt{Radzom_2024}; \citealt{Ferreira_2024}; \citealt{Wang_2024}), and is structured as follows. In Section \ref{sec:obs}, we outline our photometric and spectroscopic observations. In Section \ref{sec:stellar}, we describe our determination of stellar parameters. In Section \ref{sec:obliq_model}, we detail our modeling of TOI-5143\,c's Doppler shadow RM signal and subsequently, its stellar obliquity. In Section \ref{sec:discussion}, we place our findings in context and discuss their implications on the mechanisms driving spin-orbit misalignment as well as the origins of hot Jupiters.

\section{Observations}\label{sec:obs}

\subsection{TESS Photometry}
\label{subsec:transit}
We use the $\mathtt{Lightkurve}$ package \citep{lightkurve2018} to extract TOI-5143\,c's (TIC ID: 281837575) Presearch Data Conditioning Simple Aperture Photometry (PDCSAP; \citealt{Smith2012PDCSAP, Stumpe2012,Stumpe2014}) light curves from the TESS Science Processing Operations Center (SPOC; \citealt{Jenkins2016}), which contains 13 full transits of TOI-5143\,c spanning Sectors 45 (five transits), 46 (four transits), and 72 (four transits). To process the light curve data, we perform normalization and clip positive outliers. We then utilize the $\mathtt{transitleastsquares}$ \citep{tls2019} and $\mathtt{wotan}$ \citep{wotan} packages to identify and remove a partial transit of planet c and all transits of the putative planet b, which is ignored in subsequent modeling.

\subsection{Transit Spectroscopy with WIYN/NEID}
\label{subsec:neid}
We obtained in-transit spectroscopy of TOI-5143 using the High Resolution (HR) mode (resolving power of $R \approx 110,000$) on the NEID spectrograph \citep{Halverson2016,Schwab2016} on the 3.5\,m WIYN telescope at Kitt Peak National Observatory in Arizona, USA.  NEID is a highly-stabilized \citep{Stefannson2016, Robertson2019} fiber-fed spectrograph \citep{Kanodia2018, Kanodia2023} with a wavelength coverage of 380--930\,nm. On April 19, 2022, we captured 21 RV measurements in HR mode with 480-second exposures spanning 03:46--07:48 UT. These observations occurred under atmospheric conditions with a seeing range of 0.6$\arcsec$--1.5$\arcsec$ (median 0.9$\arcsec$) and an airmass range of $\mathrm{z} = 1.15-1.27$. At a wavelength of $5500 \text{\AA}$, the NEID spectrograph achieved a signal-to-noise ratio of 15 pixel$^{-1}$. 

The NEID spectra were analyzed using version 1.3.0 of the NEID Data Reduction Pipeline ($\mathtt{NEID-DRP}$)\footnote{Detailed information is available at: https://neid.ipac.caltech.edu/docs/NEID-DRP/}. We extract the Level 2 NEID spectra from the NExScI NEID Archive\footnote{\url{https://neid.ipac.caltech.edu/}}, and derive absorption line broadening profiles for all 21 RV observations as a function of phase and velocity in order to analyze the Doppler shadow of planet c's transit on the stellar disk. Specifically, we perform least-squares deconvolution \citep{Donati_1997} between the NEID spectra and the ATLAS9 synthetic nonrotating spectral template \citep{Castelli_2004} that most closely matches TOI-5143's stellar parameters (see e.g., \citealt{Zhou2018, dong22}).

\section{Stellar Properties}\label{sec:stellar}

To ascertain additional stellar parameters, such as stellar mass and radius, we utilize the MESA Isochrones \& Stellar Tracks (MIST) model \citep{Choi2016mist,Dotter2016mist}, combined with a spectral energy distribution (SED) fitting approach. We compile photometry from various catalogs, including 2MASS \citep{Cutri2003}, WISE \citep{Cutri2014AllWISE}, TESS \citep{Ricker2015}, and Gaia DR2 \citep{GaiaCollaboration2018}.
Gaussian priors based on our synthetic spectral fitting were applied to $T_{\rm eff}$ and [Fe/H], along with the parallax from Gaia DR3 \citep{GaiaCollaboration2023} and an upper limit for the $V$-band extinction, derived from the 3D dust map by $\mathtt{mwdust}$ \citep{Bovy2016}. We perform the SED fitting with the Differential Evolution Markov Chain Monte Carlo (DEMCMC) technique, integrated within $\mathtt{EXOFASTv2}$   \citep{Eastman2017,Eastman2019}, from which we obtain uncertainties for fitted parameters. The MCMC procedure was considered converged when the Gelman-Rubin diagnostic \citep[$\hat{R}$;][]{Gelman1992} fell below 1.01 and the count of independent draws surpassed 1,000, resulting in the following best-fit parameters for TOI-5143: $M_\star=0.864^{+0.040}_{-0.033}\,\msun$, $R_\star=0.852\pm0.020 \, \rsun $, $\rho_\star=1.97^{+0.18}_{-0.16} \, \mathrm{g/cm}^3$, $\logg=4.514^{+0.029}_{-0.027}$, $\teff=5183^{+59}_{-57} \, \mathrm{K}$, $[{\rm Fe/H}]=0.107\pm0.044 \,\mathrm{dex}$, and $\mathrm{Age}=7.4^{+4.4}_{-4.2}\,\mathrm{Gyr}$. Based on the results of \cite{Tayar2020}, we adopt the approximate 2.4\% systematic uncertainty floor on our estimate of $T_{\rm eff}$, which increases its uncertainty to $125\, \mathrm{K}$. Our final stellar parameters are listed in Table~\ref{tbl:parameters}; note that all are consistent with those reported in Quinn et al. (2025, in preparation) to within $\pm 2\sigma$.

\begin{table}[ht!]
\centering
\footnotesize
\caption{Median values and 68\% HDIs for relevant and fitted parameters of the \toi (\tic) system and its planet c (TOI-5143\,c). \label{tbl:parameters}}
\begin{tabular}{llcccc}
  \hline
  \hline
Parameters & Description/Units & Values \\
\hline\\\multicolumn{2}{l}{\textbf{Stellar Properties}}&\smallskip\\
\multicolumn{2}{l}{\gaia Parameters}&\smallskip\\
~~~~$\alpha_{\rm J2016}$\dotfill & RA (HH:MM:SS.ss)\dotfill & \ra\\
~~~~$\delta_{\rm J2016}$\dotfill & Dec (DD:MM:SS.ss)\dotfill & \dec\\
~~~~$\varpi$\dotfill & Parallax (mas)\dotfill& \parallax\\
~~~~$G$\dotfill & $G$ magnitude \dotfill & \Gmag\\
~~~~$G_{\mathrm{BP}}$\dotfill & $G_{\mathrm{BP}}$ magnitude \dotfill & \GBPmag\\
~~~~$G_{\mathrm{RP}}$\dotfill & $G_{\mathrm{RP}}$ magnitude \dotfill & \GRPmag\\
\\\multicolumn{2}{l}{Stellar Fit}&\smallskip\\
~~~~$M_\star$\dotfill & Stellar mass ($M_\Sun$)\dotfill & \thismstar\\
~~~~$R_\star$\dotfill & Stellar radius ($R_\Sun$)\dotfill & \thisrstar\\
~~~~$\rho_\star$\dotfill & Stellar density ($\mathrm{g/cm}^3$)\dotfill & \thisrhostar\\
~~~~$\log{g}$\dotfill & Stellar surface gravity (cgs)\dotfill & \thislogg\\
~~~~$T_{\rm eff}$\dotfill & Stellar effective temperature (K)\dotfill & \thisteff\\
~~~~$[{\rm Fe/H}]$\dotfill & Stellar metallicity (dex)\dotfill & \thisfeh\\
~~~~Age\dotfill & Stellar age (Gyr)\dotfill & \thisage\\

\\\hline\\
\multicolumn{2}{l}{\textbf{Doppler Shadow Fit}}\smallskip\\
~~~~$P$ & Orbital period (days)\dotfill &  \DTperiodc \\
~~~~$T_C$ & Reference mid-transit time\dotfill &  \DTmidtc\\
~~~~$\rho_{\star, \textrm{circ}}$ & Stellar density ($\mathrm{g/cm}^3$)\dotfill &  \DTrhostar\\
~~~~$b$\dotfill & Impact parameter \dotfill & \DTb \\
~~~~$R_{\rm p}/R_\star$\dotfill & Planet-star radius ratio\dotfill & \DTrprs\\
~~~~$\lambda$\dotfill & Projected stellar obliquity ($\degr$)\dotfill & \DTlam \\
~~~~$v\sin i_\star$\dotfill & Projected line broadening ($\mathrm{km/s}$)\dotfill & \DTvsini\\
~~~~$v_\mathrm{macro}$\dotfill & Macroturbulent velocity ($\mathrm{km/s}$)\dotfill & \DTnonrotv\\
\\\multicolumn{2}{l}{Derived Parameters}&\smallskip\\
~~~~$a/R_\star$\dotfill & Planet-star separation\dotfill & \DTaor \\
~~~~$a$\dotfill & Semi-major axis (AU)\dotfill & \DTap \\
~~~~$R_{\rm p}$\dotfill & Planet radius ($R_\mathrm{J}$)\dotfill & \DTrp\\
~~~~$i$\dotfill & Orbital inclination ($\degr$)\dotfill & \DTincl\\
\\\hline
\\\multicolumn{2}{l}{\textbf{Linear TTV Fit}}&\smallskip\\
~~~~$P$\dotfill & Orbital period (days)\dotfill & \ttvper\\
~~~~$T(0)$\dotfill & Optimal mid-transit time \dotfill & \ttvmidt\\
\smallskip\\
\hline
\end{tabular}
\tablecomments{\gaia parameters are obtained from the \gaia Data Release\,3 \citep{GaiaDR3} and all other stellar properties are derived in Sections~\ref{sec:stellar} and \ref{sec:obliq_model}. Mid-transit times are reported in units of $\mathrm{BJD}-2457000$. TTV parameters are derived from an MCMC fit of the 13 observed TESS mid-transit times assuming a linear ephemeris (see Appendix \ref{app:ttvs}).}
\end{table}

\section{Stellar Obliquity Modeling} 

\label{sec:obliq_model}

We perform a joint fit of the TESS photometry and our NEID RM data for TOI-5143\,c in order to derive its sky-projected spin-orbit angle $\lambda$. As revealed by a preliminary fit of the TESS data only, planet c does not exhibit significant transit timing variations (TTVs), so we directly model its orbital period $P$ and reference mid-transit time $T_C$ in our global RM fit. A more in-depth characterization of TOI-5143 c's TTVs is described in Appendix \ref{app:ttvs}.

We simultaneously model the planet's transits and Doppler shadow, i.e., the spectral distortion caused by its transit (see Section \ref{subsec:neid}), using the Bayesian inference framework implemented within the $\mathtt{exoplanet}$ package \citep{exoplanet:exoplanet,exoplanet:joss}, which is powered by $\mathtt{PyMC}$ \citep{pymc}. To reduce the computational cost of our Doppler shadow fit, we trim the TESS data to 16-hr segments roughly centered on each of TOI-5143\,c's 13 full transits (such that each segment spans $\sim 10\times$ the transit duration), which are constructed based on the linear transit ephemerides from our preliminary transit fit. We apply uniform priors on the $P$,  $T_C$, and the mid-transit time of the NEID RM observations $T_\mathrm{RM}$. To account for photometric stellar variability, we adopt a Matern 3/2 Gaussian Process (GP) kernel \citep{Matern_2013} with variance ($\mathrm{GP}_s$), standard deviation ($\mathrm{GP}_\sigma$), and length-scale ($\mathrm{GP}_\rho$) hyperparameters, which we re-parameterize to their logarithmic forms and fit for each transit.

To measure TOI-5143\,c's spin-orbit configuration, we compute a time series of the planet's position $x_{\rm p}(t)$ on the stellar disk (assuming rigid rotation) and the associated subplanetary velocities: 
\begin{equation}
    v_{\rm p}(t)=x_{\rm p}(t) v\sin i_\star,
\end{equation}
where $t$ is the time of each NEID measurement and $v\sin i_\star$ is the projected stellar rotational velocity (i.e., line broadening), which we initialize with uniform priors over [0.1, 50]\,$\mathrm{km}/\mathrm{s}$. We then determine the stellar velocity channels being occulted by the planet's shadow, modeling its velocity profile as a Gaussian with a width of $\sigma=\sqrt{v_0^2+v^2_{\mathrm{macro}}}$, where $v_0$ is a constant determined by NEID's resolution ($R=110,000$) and $v_{\mathrm{macro}}$ is the star's macroturbulent velocity, for which we adopt the same priors as $v\sin i_\star$. We also model the inferred photometric transit light curve, allowing us to scale $v_{\rm p}(t)$ by the inferred flux time series as well as the ratio of summed velocity profiles from the star and planet. Finally, the Doppler shadow likelihoods are combined to infer the projected stellar obliquity $\lambda$, which is initialized at $0\degree$ with uniform priors over [$-180$, $180$]$\degree$.

In addition to $P$, $T_C$,  $T_\mathrm{RM}$, and $\lambda$, we allow the following planetary parameters to vary freely in our fit: stellar density assuming a circular orbit $\rho_{\star,\mathrm{circ}}$, impact parameter $b$, and planet-to-star radius ratio $R_{\rm p}/R_\star$. Broad uniform priors are adopted for $b$ and $R_{\rm p}/R_\star$ to accommodate TOI-5143\,c's grazing configuration (though we impose an upper limit of 0.2 on $R_{\rm p}/R_\star$ due to the confirmed planetary nature of TOI-5143\,c), while Gaussian priors based on our SED fitting results (Section \ref{sec:stellar}) are used for $\rho_{\star,\mathrm{circ}}$. As is the case for our GP hyperparameters, we re-parameterize $v\sin i_\star$, $v_\mathrm{macro}$, $\rho_{\star,\mathrm{circ}}$, and $R_{\rm p}/R_*$ to their logarithmic forms. For both TESS and NEID, we apply a quadratic limb darkening model with limb-darkening coefficients \{$u_0, u_1$\} re-parameterized as in \cite{exoplanet:kipping13}, which are initialized at 0.3 and 0.2, respectively, for both instruments.

We optimize the sampler using the built-in L-BFGS-B algorithm \citep{ZhuOptimize} and sample the parameter posterior distributions using the gradient-based MCMC No-U-Turn Sampler \citep{Hoffman2011}. In particular, we run 4 MCMC chains with 10,000 tuning iterations, 5,000 sample draws, and a target acceptance rate of 0.95. We verify convergence via the $\hat R$ statistic, which is $\leq 1.001$ for all parameters except $b$ ($\hat R =1.029$; largely due to the TOI-5143\,c's grazing orbit) for all best-fit values. We discuss the posterior distributions and their covariances in Appendix \ref{app:correlations}.

\begin{figure*}
    \centering
    \includegraphics[width=1\textwidth]{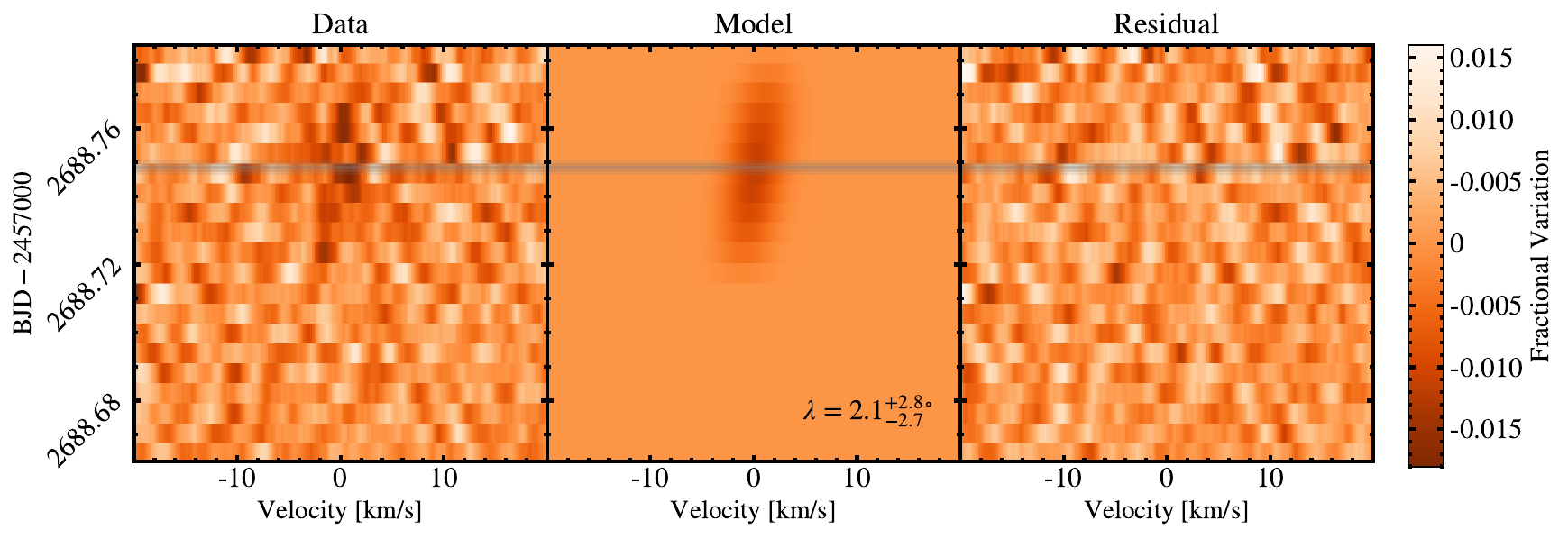}
    \caption{The Doppler shadow of TOI-5143\,c during its transit, with velocity on the horizontal axes and time on the vertical axis. The left, middle, and right panels correspond to the data extracted from the NEID spectra, our best-fit Doppler shadow model, and the residuals, respectively. The fractional flux variation of the velocity channel is represented on the color axis, and our best-fit transit midtime is displayed in each panel as the gray dashed line bounded by the 68\% ($1\sigma$) and 95\% ($2\sigma$) HDI regions.}
    \label{fig:dt}
\end{figure*}

In Table \ref{tbl:parameters}, we report the best-fit results for each parameter as the median value and their uncertainties as the 68\% highest density intervals (HDIs). Notably, as depicted in Figure \ref{fig:dt}, we find that TOI-5143\,c is spin-orbit aligned with a best-fit $\lambda=2.1 ^{+2.8}_{-2.7} \degree$. We additionally verify that all fitted parameters are in good agreement ($\leq 2\sigma$ discrepant) with those reported in Quinn et al. (2025, in preparation).

\section{Discussion} \label{sec:discussion}
TOI-5143\,c is the \emph{third} hot Jupiter in a compact, multi-planet, single-star system\footnote{There is one known hot Jupiter in a compact multi-planet system embedded within a multi-star system: TOI-942A b \citep{Wirth2021}.} to have its stellar obliquity measured (see also WASP-47\,b; \citealt{Sanchis-Ojeda2015} and WASP-84\,b; \citealt{Anderson2015}), where we define ``compact'' as having a small period ratio with a neighboring planetary companion: $P_2/P_1<6$ (see e.g., \citealt{Wang2022}). The low stellar obliquity of TOI-5143\,c ($\lambda=2.1 ^{+2.8}_{-2.7} \degree$) supports the preliminary pattern of alignment seen for hot Jupiters in compact systems. The alignment of these hot Jupiter systems, in combination with a steadily growing census of hot Jupiters hosting nearby companions (e.g., see also Kepler-730; \citealt{Canas2019}, TOI-1130; \citealt{Huang2020}, WASP-132 \citealt{Hord2022}, TOI-2000; \citealt{Sha2023}, TOI-1408; \citealt{Korth2024}, and TOI-2494; \citealt{Guerrero2021}, Quinn et al. 2025, in preparation) and updated companion rates from TTV searches \citep{Wu2023}, suggest that a non-negligible fraction of hot Jupiters arrive at their current orbits relatively quiescently, rather than through violent high-eccentricity migration pathways.

\begin{figure*}
    \centering
    \includegraphics[width=0.75\textwidth]{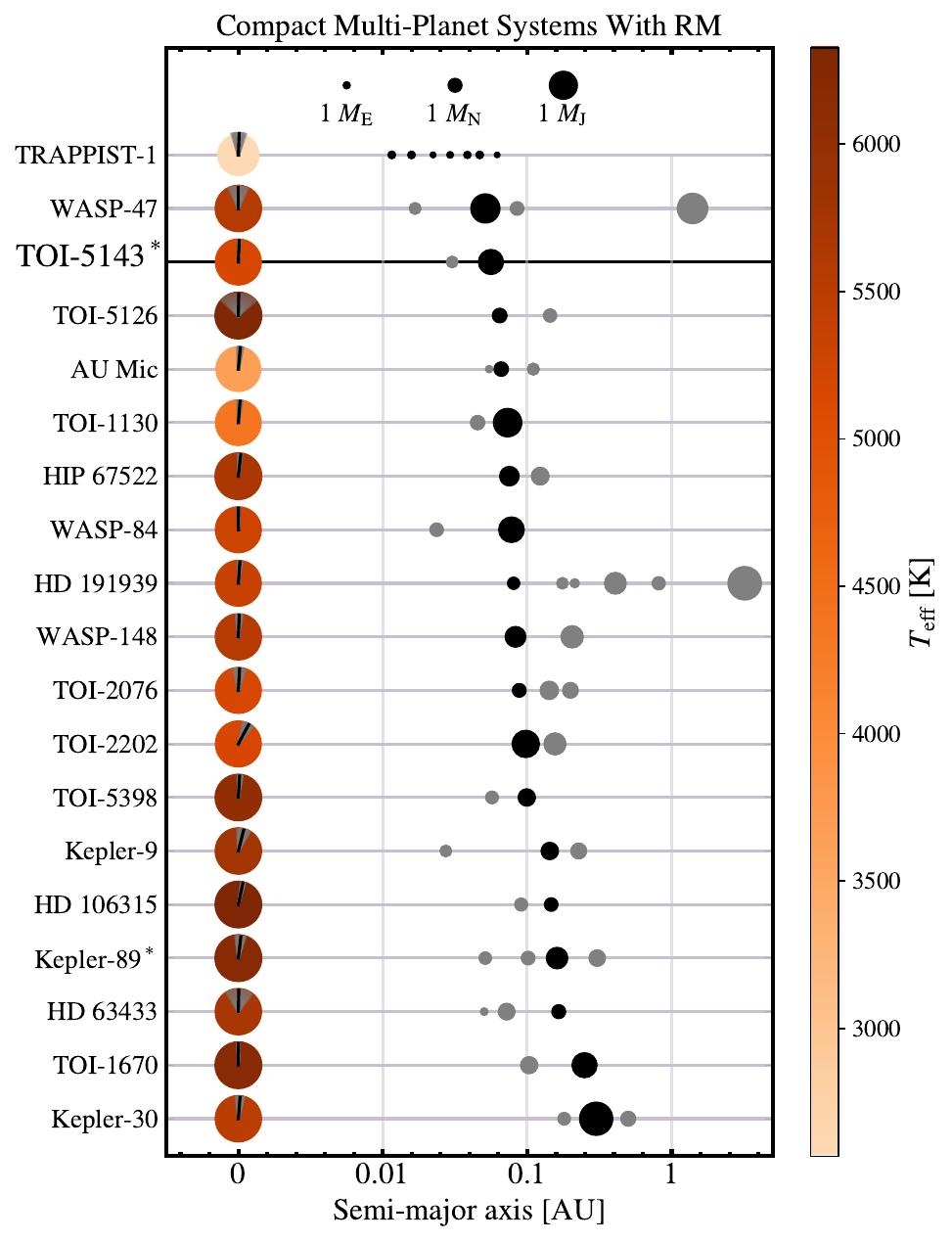}
    \caption{Visual depiction of the configurations and sky-projected spin-orbit angle for all compact multi-planet systems with an RM measurement, laterally stacked by the orbital period of the planet for which the RM effect was measured. A vertical upward spin axis indicates $\lambda=0\degree $, and planets with stellar obliquity constraints are shown in black while all others are shown in gray. Stellar effective temperatures are represented by color (see right-hand color axis). Circle sizes scale with the square root of reported planet masses to optimize visual clarity (note that for the unconfirmed planet TOI-5143\,b, we adopt the \citealt{Otegi_2020} mass-radius relation to determine mass), and the corresponding sizes of the Earth, Neptune, and Jupiter are illustrated at the top for reference. Asterisks ($^*$) indicate systems that may plausibly be binaries. Sample selection is described in Section \ref{sec:discussion}.}
    \label{fig:sys_configs}
\end{figure*}

More broadly, the aligned spin-orbit angle of TOI-5143\,c continues the trend of alignment seen across other types of exoplanets in compact systems (e.g., sub-Neptunes; \citealt{Albrecht2013} and sub-Saturns; \citealt{Radzom_2024}). In particular, combining the catalogs of \cite{Knudstrup_2024}, \cite{Albrecht2022}, and TEPCat\footnote{\url{https://www.astro.keele.ac.uk/jkt/tepcat/obliquity.html}} (\citealt{Southworth2011}, accessed on October 8, 2024), we find there are now 19 compact multi-planet single-star systems with robust RM measurements. In brief, we produce this sample by considering all secure and uncontested RM measurements for compact multi-planet systems\footnote{We consider ``compact multi-planet systems'' to host at least one planet pair with $P_2/P_1<6$ if the pair contains a Jovian-mass planet ($\geq0.3\,M_\mathrm{J}$), or $P_2/P_1<4$ otherwise (e.g., \citealt{Radzom_2024}). We note that the resultant sample is not strongly sensitive to our choice of period ratio cut.} featured in these catalogs, prioritizing values reported in \cite{Knudstrup_2024}, followed by those from \cite{Albrecht2022}, and lastly the ``preferred" values from TEPCat. We then utilize the confirmed planets catalog on the NASA Exoplanet Archive accessed on October 9, 2024, and the multi-star catalog of \cite{Rice2024} (which is based on Gaia DR3 data), to filter out systems with confirmed stellar companions. None of these 19 systems show evidence of misalignment, and \emph{all} are statistically consistent with alignment ($<1\sigma$ deviation from $|\lambda|=10\degree$ or $<2\sigma$ deviation from $0\degree$). We display the configurations of these systems in Figure \ref{fig:sys_configs}.

It is worth noting that, by virtue of their constituent planets being transit-detected, the 19 compact multi-planet systems considered here are $\sim$co-planar. Absent from our sample are systems containing one or more non-transiting planets in compact but mutually inclined orbits. Yet, TTV and transit duration variation searches indicate that the occurrence rate of such inclined planets is non-negligible \citep{Xie2014, Holczer2016, Ofir2018, ZhuPetrovich2018, Millholland_2021, Wu2023}, implying there is an unaccounted-for sample of compact multi-planet systems that are spin-orbit misaligned. However, this possibility has little effect on our conclusion of primordial alignment, as such mutually-inclined compact systems would be dynamically hotter than those consisting of co-planar planets only \citep{ZhuPetrovich2018}, and hence they would not represent pristine tracers of the primordial disk plane.

Additional evidence for primordial alignment follows from current constraints on stellar ages. While few RM measurements exist for young ($\lesssim100 \, \mathrm{Myr}$ old) planetary systems, all of such single-star systems, some still embedded in debris disks, are found to be aligned (e.g., AU Mic; \citealt{Hirano2020b}, V1298 Tau; \citealt{Johnson2022}, and K2-33; \citealt{Hirano_2024}). Further, it has been demonstrated that misaligned hot Jupiter systems tend to be older than those that are aligned \citep{Hamer2022}, and warm Jupiters in single-star systems, which often host nearby companions \citep{Huang2016, Wu2023}, tend to be aligned \citep{Rice2022b,Wang_2024}. Therefore, we find that the current census of RM measurements indicates that single-star exoplanetary systems are likely primordially aligned and may become misaligned during the post-disk phase of evolution.

Nearly all known compact multi-planet systems orbiting stars with confirmed or suspected stellar companions are aligned as well (e.g., Kepler-89; \citealt{Hirano2012}, Kepler-25; \citealt{Albrecht2013}, TOI-942; \citealt{Wirth2021, Teng_2024}, V1298 Tau; \citealt{Johnson2022}, HD 110067; \citealt{Zak2024}, and potentially HD 148193; \citealt{Knudstrup_2024}). While unconfirmed, Quinn et al. (2025, in preparation) found that TOI-5143 may have a bound M-dwarf companion at a projected separation of $116\,\mathrm{AU}$, and thus may instead constitute another such aligned compact multi-planet system within a binary. K2-290A, a hot star ($\teff\approx 6300\,\mathrm{K}$) in a triple-star system, represents one major exception, however, hosting two planets (a warm Jupiter and inner sub-Neptune) on compact and retrograde orbits \citep{Hjorth_2019,Hjorth2021}. It is known that the presence of stellar companions introduces additional pathways to excite stellar obliquity (\citealt{Batygin2012binary, Lai2014, Best_2022, Albrecht2022}), but also that planets orbiting stars hotter than the Kraft Break ($\teff \gtrsim 6250\,\mathrm{K}$; \citealt{Kraft1967}) are more often spin-orbit misaligned than those around cool stars (though this trend appears to apply mostly to hot Jupiters; \citealt{Winn2010a, Albrecht2012}). Very few RM measurements exist for compact multi-planet systems around hot stars, so it is not yet clear whether the anomalously misaligned configuration of these planets can be attributed to K2-290A's hot temperature or the presence of its stellar companions (or both).

Non-RM measurements of stellar obliquity introduce further ambiguity. For example, \cite{Huber_2013} employed an asteroseismic technique to determine that the co-planar compact multi-planet system orbiting the formerly hot, now-evolved star Kepler-56, was misaligned ($\lambda\approx 45\degree$). More recently, \cite{Louden_2024} performed a statistical study on the $v\sin i_\star$ of planet-hosting systems, finding that compact multi-planet systems may commonly be spin-orbit misaligned, especially around hot stars. More robust RM measurements of compact multi-planet systems across a wide range of planetary and stellar types, including multi-star systems, are needed to fully characterize the prevalence of primordial alignment.

\section*{Acknowledgments}
We thank the referee, Jason W. Barnes, for their constructive feedback which helped to improve the manuscript.

Data presented were obtained by the NEID spectrograph built by Penn State University and operated at the WIYN Observatory by NOIRLab, under the NN-EXPLORE partnership of the National Aeronautics and Space Administration and the National Science Foundation. These results are based on observations obtained with NEID on the WIYN 3.5m Telescope at Kitt Peak National Observatory (PI: Jiayin Dong, NOIRLab 2022A-413894). WIYN is a joint facility of the University of Wisconsin–Madison, Indiana University, NSF's NOIRLab, the Pennsylvania State University, Purdue University, University of California, Irvine, and the University of Missouri. The authors are honored to be permitted to conduct astronomical research on Iolkam Du'ag (Kitt Peak), a mountain with particular significance to the Tohono O'odham.

We acknowledge the use of public TESS data from pipelines at the TESS Science Office and at the TESS Science Processing Operations Center. All TESS data used in this paper can be found at MAST: \dataset[10.17909/2kqk-j472]{http://dx.doi.org/10.17909/2kqk-j472}. This research made use of Lightkurve, a Python package for Kepler and TESS data analysis \citep{lightkurve2018}, as well as the NASA Exoplanet Archive (\citealt{PSCompPars}, Composite Planet Data Table: \dataset[10.26133/NEA2]{https://doi.org/10.26133/NEA2}), which is operated by the California Institute of Technology, under contract with the National Aeronautics and Space Administration under the Exoplanet Exploration Program. 

M.R. acknowledges support from Heising-Simons Foundation grant $\#$2023-4478 and the National Geographic Society. S.W. acknowledges support from Heising-Simons Foundation grant $\#$2023-4050.
We acknowledge support from the NASA Exoplanets Research Program NNH23ZDA001N-XRP (grant $\#$80NSSC24K0153).
This research was supported in part by Lilly Endowment, Inc., through its support for the Indiana University Pervasive Technology Institute.

The Flatiron Institute is a division of the Simons foundation.

\vspace{5mm}
\facilities{TESS, WIYN/NEID, \emph{Gaia}, NASA Exoplanet Archive}

\software{$\mathtt{ArviZ}$ \citep{arviz_2019}, $\mathtt{astropy}$ \citep{exoplanet:astropy13, exoplanet:astropy18}, $\mathtt{celerite2}$ \citep{exoplanet:foremanmackey17, exoplanet:foremanmackey18}, $\mathtt{corner}$ \citep{corner}, $\mathtt{EXOFASTv2}$   \citep{Eastman2017,Eastman2019}, $\mathtt{exoplanet}$ \citep{exoplanet:joss, exoplanet:exoplanet}, $\mathtt{Jupyter}$ \citep{Jupyter}, $\mathtt{Matplotlib}$ \citep{Matplotlib07, Matplotlib16}, $\mathtt{NumPy}$ \citep{NumPy11, NumPy20}, $\mathtt{pandas}$ \citep{mckinney-proc-scipy-2010, reback2020pandas}, $\mathtt{PyMC}$ \citep{pymc}, $\mathtt{SciPy}$ \citep{2020SciPy-NMeth}, $\mathtt{transitleastsquares}$ \citep{tls2019}, $\mathtt{wotan}$ \citep{wotan}}

\clearpage

\appendix

\section{Transit-only Fit and TTV Analysis} \label{app:ttvs}
Quinn et al. (2025, in preparation) did not identify any strong TTVs in their analysis of TOI-5143\,c, which included TESS Sectors 45 and 46. However, our inclusion of TESS Sector 72 data (acquired $\sim2$ years following Sector 46), as well as our in-transit RM measurement with NEID, enables a substantially longer baseline to be probed. As such, we re-analyze planet c's TTVs using the best-fit orbital period and reference mid-transit time from our global Doppler shadow fit (Section \ref{sec:obliq_model}). In particular, we perform a separate transit-only fit with the orbital period fixed (to $P=5.2097118\,\mathrm{d}$), which allows us to build the Keplerian orbit for planet c and independently model each of its mid-transit times $T_{0..12}$ and hence identify potential TTVs. We otherwise adopt the same set-up with $\mathtt{PyMC}$ and $\mathtt{exoplanet}$, fitting for $\rho_{\star, \textrm{circ}}$, $b$, $R_{\rm p}/R_\star$ and TESS quadratic limb darkening coefficients $u_0$ and $u_1$. As was the case for our Doppler shadow fit, all fitted parameters obey $\hat R \leq1.001$ except $b$, indicating overall convergence. We find $\rho_{\star, \textrm{circ}}=2.03\pm 0.14 \,\mathrm{g/cm}^3$, $b=0.98\pm 0.03$, and $R_{\rm p}/R_\star =0.163\pm 0.022$, and report our best-fit observed mid-transit times in Table \ref{tbl:parameters_transit}. We verify that each parameter is consistent (within $\pm2 \sigma$) with the results of both our Doppler shadow fit and Quinn et al. (2025, in preparation) 
\begin{table}[ht!]
\centering
\footnotesize
\caption{Median values and 68\% HDIs for the observed TESS mid-transit times of TOI-5143\,c, obtained from our transit-only fit. \label{tbl:parameters_transit}}
\begin{tabular}{ccc}
  \hline
  \hline
Mid-transit Time & Value \\
\hline \\

~~~~$T_0$\dotfill & \transitmidtczero \\
~~~~$T_1$\dotfill  & \transitmidtcone \\
~~~~$T_2$\dotfill & \transitmidtctwo \\
~~~~$T_3$\dotfill & \transitmidtcthree \\
~~~~$T_4$\dotfill & \transitmidtcfour \\
~~~~$T_5$\dotfill & \transitmidtcfive \\
~~~~$T_6$\dotfill & \transitmidtcsix \\
~~~~$T_7$\dotfill & \transitmidtcseven \\
~~~~$T_8$\dotfill & \transitmidtceight \\
~~~~$T_9$\dotfill & \transitmidtcnine \\
~~~~$T_{10}$\dotfill & \transitmidtcten \\
~~~~$T_{11}$\dotfill  & \transitmidtceleven \\
~~~~$T_{12}$\dotfill  & \transitmidtctwelve \\
\smallskip\\

\hline
\end{tabular}
\tablecomments{The orbital period was fixed to the best-fit value from our global Doppler shadow fit (see Section \ref{sec:obliq_model}). All mid-transit times are reported in units of $\mathrm{BJD}-2457000$.}
\end{table}

To investigate the TTVs of TOI-5143\,c's orbit, we utilize an MCMC method with the $\mathtt{emcee}$ package to fit a linear ephemeris to the 13 observed TESS mid-transit times and compare the observed values to the resultant linear prediction. Specifically, we optimize the equation:
\begin{equation}
    T(N)= N\times P + T(0) 
\end{equation}
where $T(0)$ is the optimal zero-epoch mid-transit time, $P$ is the optimal period, $N$ is the transit epoch number, and $T(N)$ is the time at epoch $N$. This MCMC approach allows us to accurately characterize uncertainties on the expected orbital ephemeris and better identify potential deviations. Note that we select $T(0)$ to be the epoch that minimizes the covariance between $T(0)$ and $P$ \citep{Shporer_2009}. 

\begin{figure}
    \centering
    \includegraphics[width=0.7\textwidth]{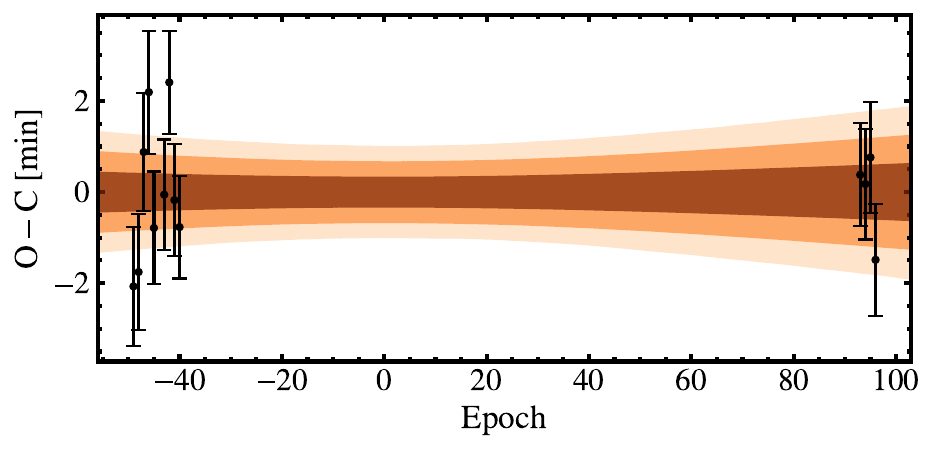}
    \caption{The difference between the observed and calculated mid-transit times assuming a linear ephemeris as a function of transit epoch. Regions containing $\pm1\sigma$ (68.3\%), $\pm2\sigma$ (95.4\%), and $\pm3\sigma$ (99.7\%) confidence intervals are overplotted in red, orange, and beige, respectively. All TTVs show $<2\sigma$ deviation from our predicted linear ephemeris.}
    \label{fig:ttv}
\end{figure}
We report the resulting orbital period and optimal mid-transit time in Table \ref{tbl:parameters}, and display differences between the observed and predicted mid-times in Figure \ref{fig:ttv}. We find no compelling evidence for TTVs--- all residuals are consistent with the expected linear ephemeris to within $2\sigma$ and the TTVs exhibit no clear sinusoidal signal. This is expected given that the two planets are relatively far from 2:1 period commensurability, and that TTV amplitude scales with orbital period and mass of the perturbing planet \citep{Holman2005, Wu2023}, which are both small in the case of TOI-5143\,c and b, respectively.

\section{Parameter Correlations From The Stellar Obliquity Fit}
\label{app:correlations}
\begin{figure*}
    \centering
    \includegraphics[width=0.9\linewidth]{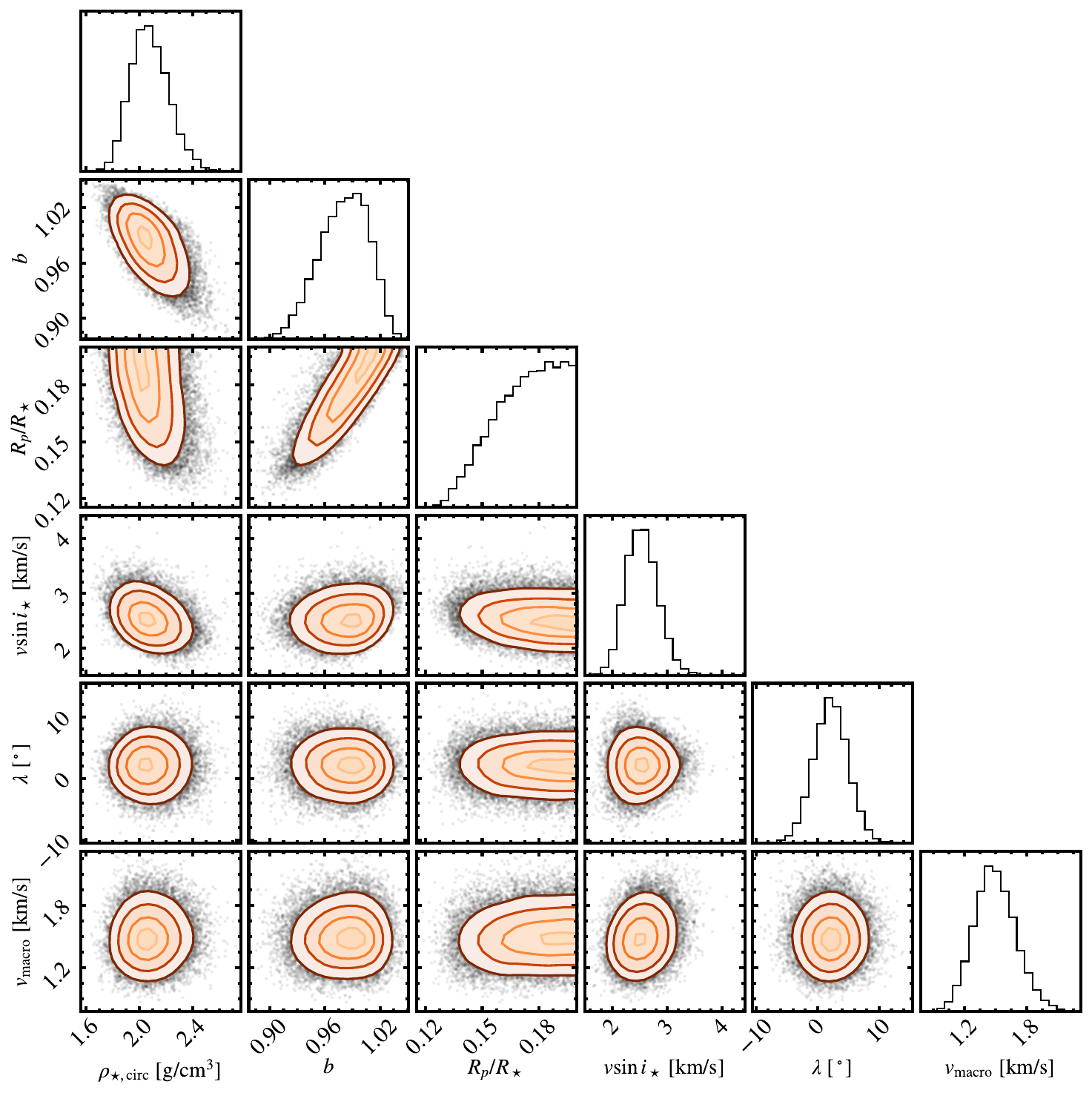}
    \caption{Corner plot of the posteriors from our Doppler shadow modeling. Yellow, orange, red, and maroon contours correspond to the 11.8\%, 39.3\%, 67.5\%, and 86.4\% HDIs, respectively.}
    \label{fig:corner}
\end{figure*}

Figure~\ref{fig:corner} displays a corner plot showing the covariance between the stellar and planetary parameter posteriors that resulted from our joint Doppler shadow model (Section \ref{sec:obliq_model}). 
The grazing orbit ($b=0.98\pm 0.03$) and unconstrained eccentricity of TOI-5143\,c lead to a degeneracy between $\rho_{\star,\mathrm{circ}}$, $b$, and $R_p/R_\star$. However, $\lambda$ and $v\sin i_\star$ show no obvious correlation, indicating that our measurement of planet c's spin-orbit angle is robust. 
Additionally, the strong agreement between our best-fit parameters (especially stellar density) and those of Quinn et al. (2025, in preparation), in which TOI-5143\,c's eccentricity was allowed to vary, validate our assumption of a circular orbit for TOI-5143\,c (note that Quinn et al. (2025, in preparation) compute a low eccentricity of $e=0.07^{+0.034}_{-0.033}$ in their global fit, which is consistent with a circular orbit to within $2.1\sigma$). 

\bibliography{dt5143}{}
\bibliographystyle{aasjournal}

\end{document}